# Accurate high-resolution depth profiling of magnetron sputtered transition metal alloy films containing light species: A multi-method approach

M. V. Moro[1, a], R. Holeňák[1, 2], L. Zendejas Medina[3], U. Jansson[3] and D. Primetzhofer[1]

[1]Department of Physics and Astronomy, Uppsala University, Box 516, S-751 20 Uppsala, Sweden
[2]Department of Physical Engineering, Brno University of Technology, 616 69, Brno, Czech Republic
[3]Department of Chemistry, Uppsala University, Box 538, S-751 20 Uppsala, Sweden

**Abstract**

We present an assessment of a multi-method approach based on ion beam analysis to obtain high-resolution depth profiles of the total chemical composition of complex alloy systems. As a model system we employ an alloy based on several transition metals and containing light species. Samples have been investigated by a number of different ion-beam based techniques, i.e., Rutherford Backscattering Spectrometry, Particle-Induced X-ray Emission, Elastic Backscattering Spectrometry and Time-of-Flight/Energy Elastic Recoil Detection Analysis. Sets of spectra obtained from these different techniques were analyzed both independently and following an iterative and self-consistent approach yielding a more accurate depth profile of the sample, including both metallic heavy constituents (Cr, Fe and Ni) as well as the rather reactive light species (C, O) in the alloy. A quantitative comparison in terms of achievable precision and accuracy is made and the limitations of the single method approach are discussed for the different techniques. The multi-method approach is shown to yield significantly improved and accurate information on stoichiometry, depth distribution, and thickness of the alloy with the improvements being decisive for a detailed correlation of composition to the material properties such as corrosion strength. The study also shows the increased relative importance of experimental statistics for the achievable accuracy in the multi-method approach.

[a] Corresponding author:
marcos.moro@physics.uu.se

## 1. Introduction

Ion beam-based analytical techniques represent a powerful set of tools for non-destructive, standard-less, depth-resolved and highly accurate elemental composition analysis in the depth regime from several nm up to few µm [1]. By changing type of incident ion, the geometry of experiment, particle energy, or by acquiring different products originating from ion-solid interaction, complementary information can be extracted. However, analysis is often challenged either in terms of mass resolution - when several comparably heavy elements are present in the sample - or in terms of sensitivity - when light species are present in heavy matrixes.

Hence, typically only a combination of several ion beam-based techniques will overcome the limitations of each individual method and provides complementary information about the sample [2, 3, 4]. The most commonly employed ion beam analysis (IBA) technique is Rutherford Backscattering Spectrometry (RBS) [5], where light primary charged particles (typically $H^+$, $D^+$, $He^{+,++}$ and $Li^+$), detected after being elastically backscattered from target nuclei and inelastically decelerated by the electronic system of the target, are used for determining concentration profiles of the target constituents. The high accuracy of the method as well as high sensitivity has made it a common tool to characterize thin film deposition processes [6] or the effects of ion implantation [7]. Even though the accuracy of Rutherford scattering cross sections is found very high for typical conditions (e.g., > 99 % for 2.0 MeV $He^+$ scattered in elements with Z > 6), limitations of RBS are found in, e.g., limited detector energy resolution, i.e., the inability to uniquely identify constituents with small relative mass differences due to similar scattering kinematics [8].

Additionally, the sensitivity for light constituents in heavy matrices is rather limited. Apart from backscattered particles, one may also detect other products of elastic nuclear collisions, i.e., recoiling target species. Time-of-Flight/Energy coincidence measurements of Elastically Recoiled target particles due to irradiation with heavy primary ions with several tens MeV's (ToF-E ERDA) [9] enables mass-resolved composition depth profiling without masking of the signals of light constituents, and with comparable sensitivity for all constituents. Employing heavy primary ions like iodine, recoil cross sections in the lab system vary only by around a factor of 5 over the whole periodic table and detector efficiencies might be deviating significantly from unity only for the lightest recoil species [10]. The method is thus very suitable to quantify e.g. light electrolytes [11] or simply the concentration levels of undesired impurities [12]. Elastic Backscattering Spectrometry (EBS) [13] can also obtain complementary isotope-resolved information on light target constituents. This method is based on using elevated energies and employing resonant non-Rutherford cross sections, making use of the fact that due to the short interaction distances scattering can no longer be described by simple Coulomb interaction of point charges. This phenomenon can enhance the probability to detect backscattered particles by orders of magnitudes with respect to expectations from Rutherford cross section. Finally, X-ray detectors allow for detecting the characteristic x-ray emission due to de-excitation of the target electronic system after passage of an ion. Particle Induced X-ray Emission spectroscopy (PIXE) [14] can be used to determine the elemental concentration of nearby elements providing in parallel a signal even for trace impurities of heavier elements in the sample [15]. The latter property is due to the fact that Bremsstrahlung is effectively suppressed in comparison to electron-based x-ray excitation techniques.

As mentioned above, a combination of these techniques may be beneficial when the samples of interest contain light species in a heavy matrix, and strong gradients in concentrations may be expected [16, 17]. To qualitatively and quantitatively assess the advantage of such a combinatorial approach is of particular relevance since such complex chemical compositions are nowadays getting more abundant in many of the high-tech coatings employed today in mechanically or chemically challenging environments.

In this work, we present an iterative and self-consistent IBA analysis of carbon-containing transition metal alloys with light contaminants with a twofold goal. We critically assess the self-consistent approach adopted in this study, which combines different ion beam-based methods by simultaneously fitting experimental data where information obtained from each technique is used as a boundary condition for another. We compare the achieved accuracy to the ones obtained by the individual methods. In parallel, we show that a highly accurate full description even of complex samples of interest can be provided, which can yield improved understanding of the material properties and sample preparation pathways.

For the present study, sputtered thin films of C, Fe, Cr and Ni were selected as a challenging model system. The motivation behind studying this particular material system is their similarity to high-entropy alloys (HEAs), i.e., a new class of alloys that consist of four or more principal metallic elements at near-equimolar composition [18]. The resulting alloys often exhibit rare combinations of useful properties, such as high strength and high resistance to corrosion [19]. Due to the number of principal elements in a HEA, there is a unique opportunity to tune the material properties by adjusting the composition of the alloys [20].

To use this combinatorial method, the composition at any point in the films must be accurately determined, especially the carbon content. It is also necessary to measure the amount of oxygen contamination in the films. The combination of several metallic elements with similar atomic numbers and the presence of C and O amounts makes the quantification of the sample a true challenge.

## 2. Experimental procedure

### 2.1 Sample Preparation

The thin films were deposited in a home-built ultra-high vacuum magnetron co-sputtering system with a base pressure of $10^{-7}$ Pa at 300 °C. Argon gas at 0.4 Pa was used to ignite the plasma and the substrate was $SiO_2$ grown onto p-type Si (100) wafers. Fe, Ni, Cr and C-graphite targets (purity ≥ 99.95%) were arranged around the substrate at an angle of 39° with respect to surface normal of the substrate (see Fig. 1, panel a). The graphite target was powered by a pulsed DC source with a frequency of 100 Hz, while the remaining targets used separate, non-pulsed DC sources. The substrate holder was not rotated, thus creating a compositional gradient in the films. Before depositing the films, a thin layer of Cr was deposited in order to increase the adhesion to the substrate (see Fig. 1, panel b). Following deposition, a 1x1 $cm^2$ piece from the center of the film was selected for ion beam analysis. The composition and thickness of the layers are discussed in details in Sec. 3.

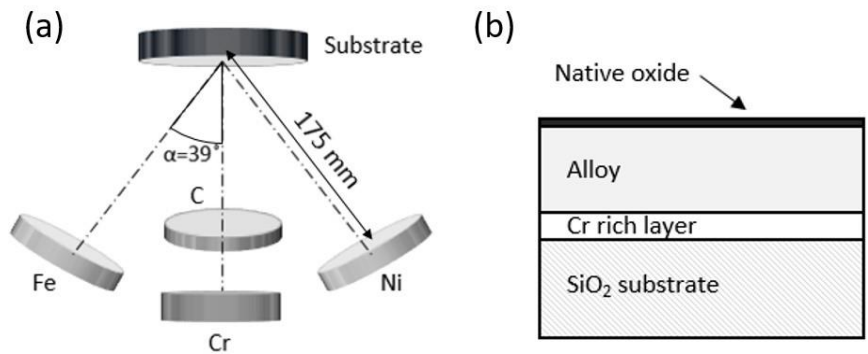

**Figure 1.** (a) Schematic illustration of the sputter chamber geometry: the four targets are positioned relative to the substrate. (b) Schematic cross-section of the sample, with the layers that are expected to be observed. The thickness of the layers is not to scale.

### 2.2 Accelerator and scattering chamber

The IBA measurements were carried out at the Tandem Laboratory at Uppsala University, using a 5-MV NEC-5SDH-2 tandem accelerator. Experiments were performed in two different chambers allowing for different techniques. The first chamber features passivated implanted planar silicon (PIPS) detectors for Rutherford Backscattering Spectroscopy (RBS), Elastic Backscattering Spectroscopy (EBS) and particle-particle Nuclear Reaction Analysis (NRA) and a silicon drift detector (SDD) for Particle-Induced X-Ray Emission (PIXE). It also holds a telescope tube for Time-of-Flight/Energy coincidence Recoil Detection Analysis (ToF-E ERDA) (see Ref. [21] for further details). The second chamber is equipped with another telescope tube for ToF-E ERDA measurements using an ionization gas chamber (GIC) (discussed below) as an energy detector as well as additional solid-state detectors for other IBA techniques. Both chambers feature sample-holders mounted on goniometers, which are remote-controlled, enabling simultaneous data acquisition and sample movement.

In the present study, RBS, EBS and PIXE measurements were performed in the first chamber, while ToF-E ERDA analysis was conducted in the second one. There are two motivations for performing ToF-E ERDA in the second chamber for the present study: at first, when heavy elements are present, the GIC does not suffer from radiation damage due to heavy recoils and scattered primaries to the same extend as a solid-state detector. Second, the mass resolution for heavier components in the second system is typically found superior compared to the first system.

### 2.3 Ion beam analysis

RBS measurements were performed using 2 MeV He$^+$ primary ions. Since the thin film was deposited onto silicon dioxide, which may show crystallinity, (e.g. in the form of texture), the goniometer was programed to randomly change the incident/exit angles of the sample in small steps (± 2°) around an equilibrium position along the measurements in order to decrease possible effects from residual channeling. Despite this effect is not affecting the signals from the thin film directly, it would influence the quantification of the charge-solid angle product from the substrate signal which can be helpful in the analysis. The PIPS detector used has a resolution of FWHM ≈ 13 keV for the whole detection chain, and it is placed at θ = 170° scattering angle, with solid angle ΔΩ = (2.16 ± 0.11) msr. RBS measurements were carried out simultaneously with PIXE, and the total charge collection (needed for more quantitative PIXE analysis) was deduced by fitting the substrate signal in the RBS spectrum. The uncertainties involved in these measurements are discussed in details in Sec. 3.2.

EBS measurements were carried out using the elastic $^{16}O(\alpha,\alpha_0)^{16}O$ resonance at 3.037 MeV He$^+$ energy, which features a scattering cross section ≈ 35 times higher than the Rutherford value [22, 23]. Since EBS spectra can be very sensitive to the specific beam energy, one can scan the projectile energy in order to depth-profile the oxygen concentration into the film [24]. For an accurate oxygen depth-profile using EBS, the accelerator beam energy was beforehand calibrated, and the beam energy is known better than 0.5%. Details on the employed energy calibration procedure for the primary beam can be found in details in Ref. [25].

ToF-ERDA enables depth-profiling the elemental composition of thin films in a depth range of ≈ 1 μm, within a depth resolution of ≈ 30 nm close to the surface. The mass-separation of the recoiled ion species is accomplished by measuring their time-of-flight and energy in coincidence (ToF-E). In this work, the ToF-E ERDA measurements were done by delivering 36 MeV $^{127}I^{8+}$ ions as probe beam and using the ToF-E telescope tube mounted in the second chamber at the fourth beam-line of the Tandem accelerator. The samples were mounted with the sample normal positioned under 67.5° with respect to the incident beam. The ToF-E telescope is fixed at 45° with respect to the direct beam. Further details on ToF-E ERDA instrumentation at Uppsala University can be found in [26]. The detection efficiency in the ToF-detector - which differs from unity in particular for light recoil species - has been corrected in the analysis code [27].

For PIXE, x-rays are detected by a 500 μm thick silicon drift detector (SDD) placed at θ = 135° with respect to the primary beam. The x-ray SDD has a resolution of FWHM ≈ 136 eV for Fe-$K_\alpha$ characteristic energy, and a solid angle of ΔΩ = (1.875 ± 0.056) msr. A 79.5 μm Mylar absorber is placed in front of the 12.5 μm Be-window of the SDD to attenuate the low-energy characteristic x-rays (e.g., from Si) and Bremsstrahlung in order to decrease the dead time of the detecting system and to protect the detector from radiation damage due to backscattered particles.

## 3. Results and discussions

### 3.1. Iterative self-consistent characterization

In Fig. 2, we show an experimental RBS spectrum (black solid-dotted line) recorded for incidence and exit angles of 5° with respect to the surface normal of the sample. In the following, we will present the results of a straightforward analysis of this dataset exclusively and compare them to results obtained performing iterative analysis using complementary methods. After the presentation of the fits and an analysis of their quality we will also discuss the individual analysis methods employed in more detail. For this aim, Fig. 2 also holds different fits obtained using the latest version of SIMNRA [28] (red solid lines in panel (a) and (b) - other colors for constituents). The stopping power data used in all the fits presented in this paper was retrieved from the most recent version of SRIM-2013 code [29]. For the fits shown in Fig. 2, the built-in fit routine of SIMNRA was used, which allows to change areal thickness and stoichiometry of a specific layer of the sample simultaneously until convergence is reached. The electronics calibration and resolution were determined beforehand (independent from the investigated sample) and kept constant during the fits. The accumulated charge of the RBS spectrum was obtained by a fitting the $SiO_2$ region of the experimental spectrum (see Fig. 2) simultaneously with the alloy layer thickness. As apparent from the fits to the experimental RBS data in Fig. 2, a broad signal with extended plateau starting at the highest energies corresponds to ions backscattered from the metallic alloy constituents (Ni-Fe-Cr).

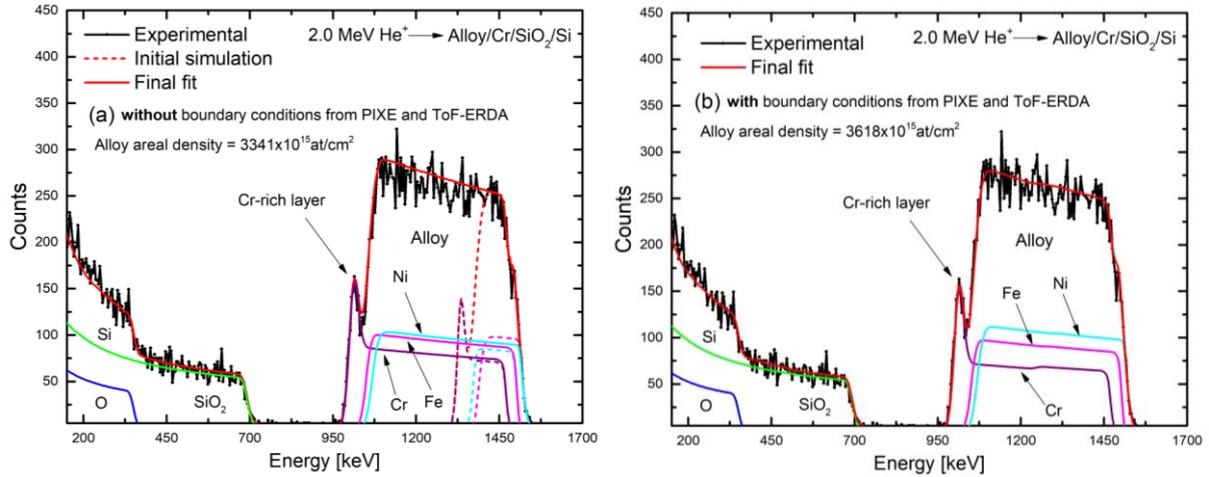

**Figure 2.** (Color online) The black solid line represents the experimental RBS spectrum recorded for 2.0 MeV He$^+$ primary ions scattered from the alloy thin film, as described in the Sec. 2. The red solid line represents the best fit provided by the SIMNRA code. Other color lines indicate the main constituents visible in the fitted spectrum. In (a), data were evaluated without further complementary input from PIXE/ToF-E ERDA results (dashed lines in (a) represent initial condition to the fit – substrate signal not shown). In (b), information from these complementary techniques was used as boundary condition to the fit (see text for details).

For the fit shown in Fig. 2 (a), we did not assume any previous knowledge regarding the sample besides what was is visible in the RBS spectrum and known from the deposition process, i.e. the presence of three metals (Cr, Fe and Ni), which have been co-deposited following a Cr seed layer on SiO$_2$-substrate. As initial condition - requested by SIMNRA to initialize the fitting - we assumed a homogeneous concentration of metallic elements in equal fractions (≈ 33.3 at.% of Cr, Fe and Ni), with an areal thickness of 1000 x 10$^{15}$at/cm$^2$. Likewise for the Cr buffer layer, we presumed an initial thickness of 100 x 10$^{15}$at/cm$^2$ (see dashed-lines in panel a). On the other hand, as boundary condition for the input fit presented in Fig. 2 (b), we used information obtained from other techniques, e.g., stoichiometry of Cr, Fe and Ni from PIXE, and carbon depth-profile from ToF-E ERDA (discussed in details below). Hence, a two-layer model was adopted to describe the carbon-containing alloy layer (as result from ToF-E ERDA, and discussed in details below): one with thickness fixed at ≈ 1750 x 10$^{15}$ at/cm$^2$, and the second kept free for the RBS fit.

The accuracy of the fits without and with boundary inputs from other IBA techniques was determined by calculating the average ratio (in %) between the fits and the experimental data $\sum(Fit/Exp)/n \times 100\%$, where *n* represents the number of channels corresponding to the range from 955 keV to 1550 keV, and it was found to be 4.3 % and 1.5 %, for the fits without and with boundary conditions (panels (a) and (b) in Fig. 2), respectively. The significant difference in fit quality can be also assessed by an analysis of the residual of the fit, as presented in Fig. 3, which shows the distribution of obtained residual normalized to the expected standard deviation. Clearly, the fit relying on automatic fitting of the RBS-spectrum is shifted from the expected normal distribution with an expectation value of -0.6.

In contrast, the multi-method fit almost perfectly coincides with expectations for a perfect fit to data with the given experimental statistics (see dashed line in Fig. 3 for comparison).

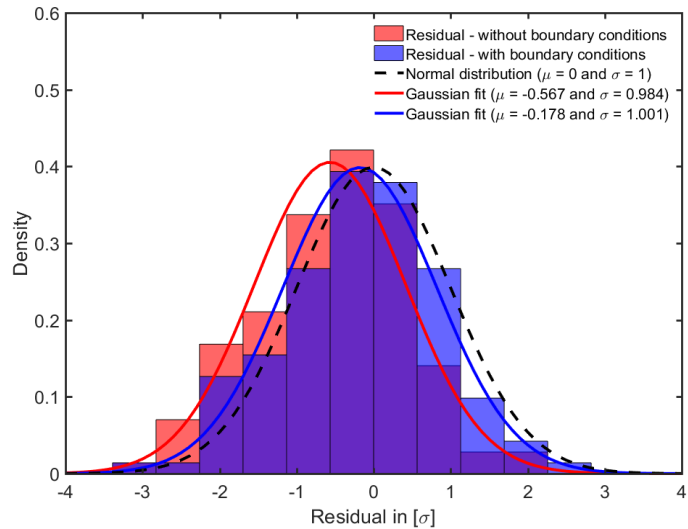

**Figure 3.** (Color online) Density distribution of the residual (presented as a fraction of the expected standard deviation - σ) between the fit and the RBS data. Red bars: fit without boundary condition, blue-bars: fit with boundary conditions from other techniques. Solid lines are Gaussian fits to the distributions. For comparison, the dashed line represents a normal distribution with mean value μ = 0 and standard deviation σ = 1.

Although a visual inspection of the fit might suggest that an apparently good fit can be obtained in both situations, the quality of the fit shown in Fig 2 (a) is thus inferior, with drastic consequences on the accuracy of the information obtained. In fact, the observed differences in thickness and concentrations between the models employed in Fig. 2 (a) and (b) are far larger than the observed difference between fit and experiment. For example, the total areal thickness of the alloy differs by around 8 %. At the same time, the accurate stoichiometry of Cr, Fe and Ni is rather difficult to be derived directly from the RBS fit, as the signals from these species strongly overlap. Discrepancies of up to 10 % for the individual constituents are possible without strongly affecting the fit quality beyond the above mentioned value. Moreover, the amount of carbon – virtually invisible in the RBS spectrum – needs to be considered during the fit to properly estimate the energy loss of the ion in the alloy layer. Note, however, that the integral areal density of the metal components, however, is obtained with higher accuracy.

While this result shows the expected advantage of employing multiple techniques putting constraints in the evaluation [30], it most importantly shows the necessity of a quantitative evaluation of the fit quality. For this purpose, SIMNRA is capable of providing chi$^2$ and the reduced-chi$^2$ for selected areas of interest in the spectrum. Based on such an assessment, the fitting has to be improved by additional constraints, which, as in the present case can be readily available from other IBA methods, and, as for the PIXE data recorded in simultaneously with RBS.

Similar improvements in deduced data due to complementary information, are also observed for the other techniques while being capable of fitting data with high quality with and without the additional boundary conditions. In Fig 4, the experimental PIXE spectrum (black dotted line) as well as the corresponding fit using the GUPIX code [31] (red solid line) are shown. The characteristic X-ray signals originating from the main metallic species present in the film are the dominant structures in both experimental and fitted data (Cr, Fe, and Ni). A signal from the Si substrate can also be distinguished. The peaks corresponding mainly to the K-shell emissions of the elements present in the alloy are well defined in the spectrum without overlaps. This data enables quantification of near-mass elements with much higher accuracy than a fit to the RBS spectrum exclusively. Additionally, by analyzing the main peaks present in Fig. 4, one can notice no evidence for heavy trace elements (Z >11 in the sample) within a quantification limit of better than ≈ 0.1 at.%, indicating a clean sample preparation routine.

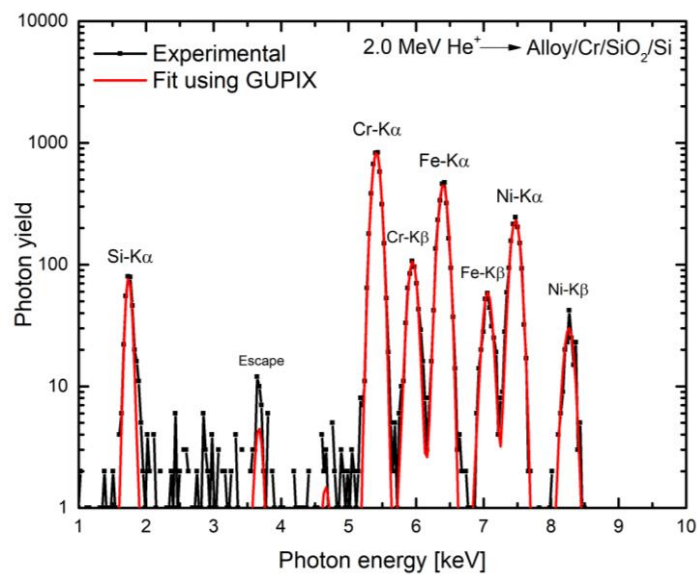

**Figure 4.** (Color online) Experimental PIXE spectrum of the alloy sample (black dotted line) recorded together with RBS. The fit provided by the GUPIX code is also shown for comparison (red solid line).

Aiming to obtain an accurate description of the stoichiometry of the metallic elements present in the alloy with high accuracy, the GUPIX code [31] was used with the integrated charge deduced from the *particles*ΔΩ* product in the RBS fit. Other fit parameters, such as the absorption filter, detector efficiency and Be-window thickness have been considered in the code. Furthermore, in the GUPIX code, the sample structure was defined as *Layer Thickness Interaction* (LTI)*, where the layer thickness is provided as input (with an initial value known from RBS and further input provide from ERD and EBS, see below). Note, that even if the fit results with and without the LTI option were rather similar a difference of around ≈ 0.8 at.% in the metallic stoichiometry was observed. For calculating input for the iterative evaluation procedure, we kept this option active, as GUPIX computes self-ionization and matrix corrections more accurately.

As final result from PIXE analysis, the total relative stoichiometry of the metallic elements in the sample is found to be Cr ≈ 31.8 at.%, Fe ≈ 33.5 at.% and Ni ≈ 34.7 at.%. The statistical uncertainties involved in their quantifications are discussed in details in Sec. 3.2. Note, that the present PIXE results, as well as the above-mentioned difference dependent on the employed model are only providing relative concentrations. These values, however, are apparently obtained with high precision. As fitting in GUPIX does not support fitting of complex multilayer systems, we have performed data analysis along different routines. Two contributions to the obtained x-ray yield can be straightforward identified to potentially require additional input on the system: both, the absorption in the material as well as the change in excitation cross section as a function of ion energy will introduce an inherent thickness dependence of both absolute intensity and the obtained relative signals of the constituents. For the present system, with the energy loss of the ions in a single transmission through the film being around 10 % of the primary energy, the changing excitation cross sections are expected to only mildly influence the signal, even in case of an extremely uneven distribution of the present element (as e.g. a buffer layer of Cr). The attenuation length of the observed x-ray energies is on the order of a few μm resulting in less than 10 % attenuation for the present film.

A GUPIX fit now yields the total concentration of all the metallic elements present in the sample, which means the final value for Cr indicates the total contribution from both alloy and buffer layer. Once from the RBS analysis the thickness and composition of the thin Cr layer can be resolved, the final relative stoichiometry of the metallic elements in the alloy could be extracted interactively with the RBS analysis, and is found to be Cr ≈ 27.9 at.%, Fe ≈ 35.8 at.% and Ni ≈ 36.3 at.%. Considering that only about 10 % of the total Cr inventory are located in the buffer layer, the effects of attenuation and reduced excitation cross sections can be estimated to affect the deduced Cr-concentration only on a sub-percent level. To evaluate this hypothesis, we made a test with GUPIX by defining a sample structure within *Fixed Matrix Solution* where the thickness of the Cr-buffer layer from RBS was also used as an input for the fit. As a result, the difference of the total amounts of Cr between the approaches was found to be only ≈ 0.3 % (relative), in accordance with our considerations. This difference, in turn, is smaller than the achievable accuracy due to experimental statistics from the present data as deduced in both PIXE and RBS.

As both RBS and in particular PIXE are rather insensitive to light species, in order to obtain absolute quantification the abundance of light elements in the alloy has to be quantified by a different approach.

For this aim, two additional IBA techniques were employed self-consistently with the others: ToF-E ERDA and EBS. In the former we depth-profiled the amount of carbon present in the film and checked for the presence of other light impurities such as hydrogen (quantification limit ≈ 0.5 at.%). In the latter we depth-profile, with higher accuracy, the amount of oxygen in the alloy.

In Fig. 5, two experimental EBS spectra (black solid line) are shown for He$^+$ projectile energies of 3.037 MeV and 3.047 MeV, panels (a) and (b), respectively. The experimental EBS spectra for each energy have been evaluated using the Multi-SIMNRA code [32]. Scattering cross sections are provided by *SigmaCalc* for the non-Rutherford resonant cross-sections [22, 33] as discussed in Sec. 2.3. In fact, the non-Rutherford $^{16}O(\alpha,\alpha_0)^{16}O$ resonance has a narrow shape (≈ 10 keV) at 3037 keV, thus we have modulated the helium beam by energy steps of the same width (corresponding to a distance of ≈ 10 nm travelled in the material - assuming bulk density). This shifting of the resonance peak allows for depth profiling the amount of O in the sample. By comparing the panels in Fig. 5, one can see a small amount of oxygen (9.23 x 10$^{15}$at/cm$^2$ ≈ 5 nm) present only at the sample surface, indicating high resistivity against corrosion. Moreover, for the final EBS fits, we have also enabled the option for corrections on *Plural and Multiple Scattering (PS and MS)*, which, in principle, improves the accuracy of the fits around the lower energy region of the EBS spectra. This correction does not affect the results in the alloy region of the film, but increase the fitting accuracy towards the SiO$_2$ layer (see dashed lines in Fig. 5).

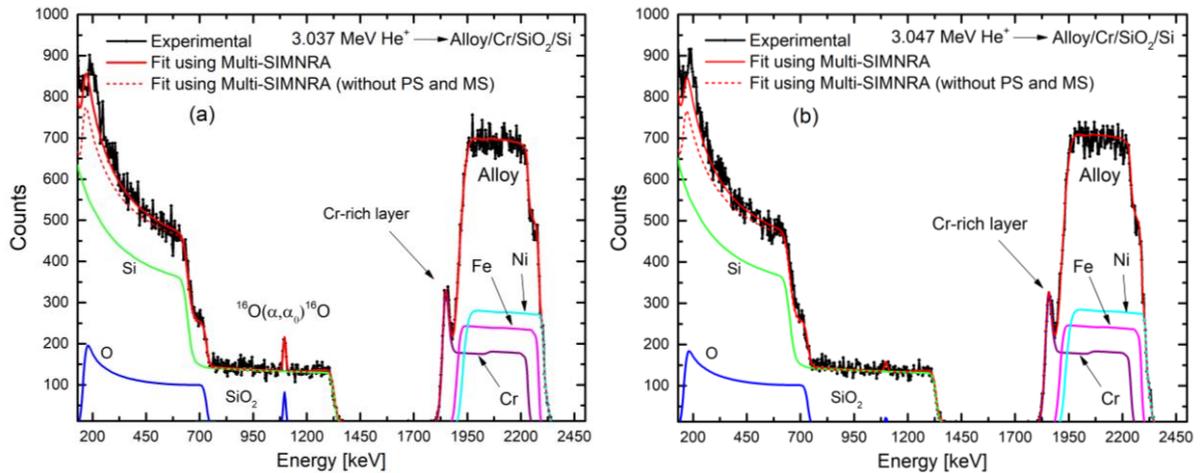

**Figure 5.** (Color online) Panel (a) Experimental EBS spectrum recorded at the resonance energy of 3.037 MeV, including the corresponding Multi-SIMNRA fit upon different energy spectra recorded during the EBS scan. Panel (b) same as in (a), but now the beam energy was 3.047 MeV. Oxygen was detected only on the surface of the alloy (see text for details). For comparison, the dashed lines represent the same fits but without corrections for Plural (PS) and Multiple (MS) scattering (see text for details).

In Fig. 6, the depth-profile of the constituents found in the alloy deduced from experimental ToF-E ERDA spectrum is shown. The depth-profile was obtained by using the POTKU code [34], and considering the efficiencies of the ToF detectors. From this figure, a rather homogeneous depth-profile for the metallic constituents of the film (Cr, Fe and Ni) ranges from the surface until a depth of

≈ 3500 x 10$^{15}$at/cm$^2$. The total areal thickness obtained estimated by the half height of the metal signal dropping at large depth is exceeding 4000 x 10$^{15}$at/cm$^2$ in contrast to RBS and EBS. This finding can be explained by two factors: first, the expected higher uncertainty of the inelastic energy loss of the heavy primary ion species as well as the recoils [35]. Second, at larger depth, the inevitably increasing contribution from nuclear energy losses equivalent to multiple small angle scattering events is deteriorating depth scales. Additionally, in Fig. 6 one can see depth-profiles of the other light elements (O, C and Si) present in the alloy (other colors). Considering the ToF-E ERDA system and its geometry, the mass resolution for heavy and nearby elements (Cr, Fe and Ni) is relatively poor; hence, their mass signals are overlapping in the mass spectrum (not shown). Here, we summed them up and indicated as "metallic alloy" (black solid line).

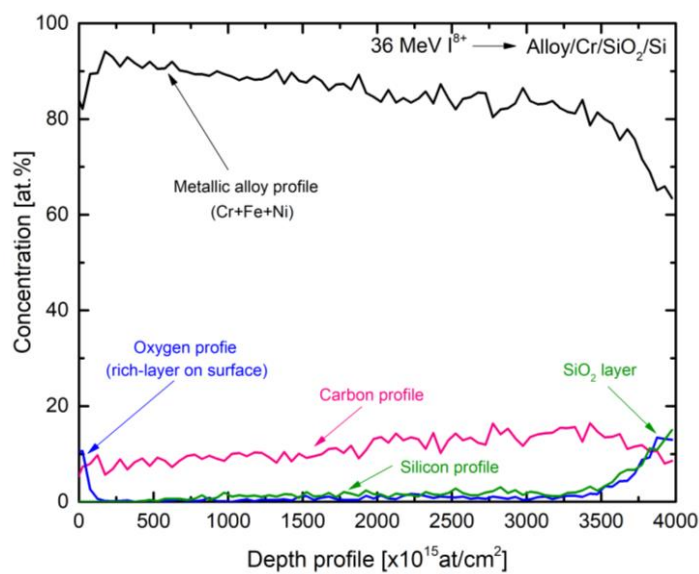

**Figure 6.** (Color online) Atomic concentration of the main constituents of the alloy as a function of their depth-profile deduced from the experimental ToF-E ERDA spectrum using the Potku code.

According to the Fig. 6, the film features a carbon concentration of ≈ 8 at.% close to the surface, slowly increasing up to ≈ 14 at.% nearly its interface to SiO$_2$ layer (blue and red lines). Note that in ToF-E ERDA, in particular for the employed heavy primary ions, the previously mentioned increasing plural and multiple scattering with decreasing energy, affects the signals. This fact, together with the associated energy loss straggling may in principle have a deteriorating effect in the obtained depth profiles, as the above mentioned effect on depth scales might differ for different recoiling species. To confirm or reject the observed gradient, a comparison with RBS is advantageous. Aiming to account for carbon in the self-consistent analysis – i.e. in the RBS and EBS fits - the carbon depth-profile was sliced into two different regions of similar thickness. In the first region ≈ [250 - 1750] x 10$^{15}$at/cm$^2$, the average carbon concentration was ≈ 9 at.%, whereas in second one, ≈ [1750 - 3250] x 10$^{15}$at/cm$^2$, a value of ≈ 12 at.% was found. These two layers with different carbon concentrations were subsequently included into the RBS and EBS fits accordingly (i.e., two-layer model). The effect of this addition can be seen as a slight depletion on the low-energy signal from the alloy peak in Figs. 2 (b) and Fig. 5, improving the overall agreement with the experimental data. Thus, on the one hand the gradient can be considered as confirmed, and, in turn, the knowledge on the depth-dependent carbon profile from ToF-ERDA, although relatively small, plays a crucial role in an accurate RBS analysis (discussed above). In Fig. 6, we can also observe the presence of a thin oxygen-rich layer on surface of the film (blue solid line). Limited energy resolution and energy-loss straggling of the recoiled atoms, together with a more complex calibration of the energy-time coincidence measurements may, however, lead to poor quantification of the thickness and concentration of oxygen direct from the depth-profile shown in Fig. 6. No detectable hydrogen was found in the film, in accordance with the corrosion-resistivity expected for such metal-blends.

### 3.2. Budget of uncertainties

The goal of the following discussion is to perform a quantitative analysis of the main sources of uncertainties affecting our self-consistent approach. A summary of the main results deduced from different ion-beam probing techniques on the alloy thin-film, together with their associated budget of uncertainties is presented in Table 1. The three main sources of statistical uncertainties shown in Table 1 were evaluated following definitions of the Joint Committee for Guides in Metrology [36], and are classified as *type A* (i.e., evaluated by methods based on statistical analysis) [37]. The total contribution of these uncertainties for each key-information regarding the sample (e.g., different layers and their constituents) [38] are also stated in the Table 1.

**Table 1.** Budget of the main sources of uncertainty affecting the depth-profile obtained by employing four different IBA techniques (RBS, PIXE, EBS and ToF-ERDA). Note that the *quantity* units are shown above their values, whereas the statistical errors are all given in %.

|  | O-layer (surface) | Alloy layer | | | | | Cr Layer | SiO$_2$ Layer |
| --- | --- | --- | --- | --- | --- | --- | --- | --- |
|  | O [$10^{15}$at/cm$^2$] | Cr [%] | Fe [%] | Ni [%] | C* [%] | Thickness [$10^{15}$ at/cm$^2$] | Thickness [Å] | SiO$_2$ [µm] |
| *Quantity, [units above]* | **9.23** | **27.8** | **35.9** | **36.3** | **10.5** | **3618** | **154** | **1.2** |
| *Counting, [%]* | 4.3 | 1.7 | 2.3 | 3.1 | 1.9 | 0.59 | 2.6 | << 1 |
| *Background**, [%]* | 2.0 | ≤ 0.25 | ≤ 0.25 | ≤ 0.25 | 1.0 | << 1 | << 1 | 3.1 |
| *Fitting precision, [%]* | 0.36 | < 1 | < 1 | < 1 | --- | 0.38 | 0.64 | 2.6 |
| **Total uncertainty (stat.), [%]** | **4.8** | **2.0** | **2.5** | **3.3** | **2.1** | **0.70** | **2.7** | **4.0** |
| *Main IBA technique* | EBS | PIXE | PIXE | PIXE | ERDA | RBS+EBS | RBS+EBS | EBS |

*\* Average over two different depth-integrated regions (see text for details).*
*\*\*Possible sources: multiple scattering (RBS/EBS/ERDA), bremsstrahlung (PIXE) and pile-up.*

As it can be seen in Table 1, the dominant source of uncertainty comes from the statistical counts of the spectra, which can be in principle improved by longer measurement time. However, we aimed to acquire all the spectra in a regime of low current to further reduce pile-up contributions, which is almost negligible for all the analysis. Contributions from plural and multiple scattering due to the backscattered particles in the heavy-elements in the metallic alloy were considered in the fits as well and belong to the background contribution category. For instance, the oxygen resonant peak is located below the heavy-element peaks (see Fig. 4, panel a), and it sits onto the Si-background, which means the uncertainty in the Si background enters the oxygen quantification accuracy.

Although not shown in Tab. 1, our results are also subject to systematic uncertainties. For RBS and EBS analysis, the major systematic uncertainties are related to the *(particle\*ΔΩ)* product [39] as obtained by fitting the signal of the substrate, due to two different causes: possible inaccuracies in the stopping power and residual channeling effects. Recent energy loss studies have demonstrated that either fully theoretical or semi-empirical stopping power models are expected to agree to experimental data within ≈ 1 % for $H^+$ and $He^+$ projectiles at energies ≥ 1.0 MeV [40, 41]. Nevertheless, in some particular cases, even the most recent tabulated stopping power values for light projectiles, as well as SRIM predictions, are found to be problematic, especially for reactive transition metals (such as vanadium) [42].

In this work, to fit the product particles times detector solid angle (i.e., *particles\*ΔΩ)* we used an energy region in RBS and EBS spectra below 1.5 MeV, corresponding to the backscattered particles from the $SiO_2$ layer. The stopping power at this energy is expected to be more accurate than the statistical uncertainties shown in Table 1; hence, we did not consider it in the budget of uncertainties. Besides our efforts to "randomize" the α-backscattering yield by rotating the sample, residual channeling due to any form of texture should be mentioned as a potential source of minor uncertainties (see for instance discussions in Ref. [43]). Since we, however, applied the same rotation procedure whenever recording any RBS or EBS spectrum, and were able to fit all spectra self-consistently, the impact of any residual channeling in a spectrum onto accuracy of the *particles\*ΔΩ* and consequently on the composition deduced is expected to be much smaller than the statistical errors in shown in Table 1.

For a stand-alone PIXE analysis, systematic sources of uncertainty *a priori* would be worse. Inaccurate *particles\*ΔΩ* values, eventual discrepancies on detector solid angle and perhaps problems with the internal GUPIX databases for x-ray production and absorption and matrix corrections, would lead in principle to higher systematic uncertainties. However, note that only a relative concentration of Cr-Fe-Ni in the sample was fitted by GUPIX, hence systematic errors is rather canceled out during the fit.

For the ToF-ERDA, systematic uncertainties related to unknown energy loss of the recoiled atoms, connected to inefficiencies of the ToF detectors (worsening as lighter as the recoil atom is) are expected to be ≈ 5-10 % of the deduced absolute concentrations in a stand-alone ToF-ERDA analysis.

To summarize, the overall systematic errors in the multi-method approach are expected to be much smaller than the observed statistical accuracies, which has important implications for planning an experimental campaign. When performing a multi-method approach, maximizing experimental statistics, which for non-destructive IBA techniques is often straightforward, gains weight in comparison to the individual approaches. There, in contrast, the described systematic uncertainties and limitations of the methods commonly set a much higher limit for the achievable accuracy.

## 4. Summary and conclusions

In this work, a high-resolution depth profiling study using different ion beam analytical techniques in an iterative and self-consistent approach to characterize a co-sputtered carbon-containing alloy thin film on silicon dioxide has been performed. The employed thin film system was chosen as a representative challenging system for such quantitative analysis as it can be considered as a model system for the emerging class of multi-functional high-entropy materials.

A qualitative and quantitative comparison of stand-alone analysis using the individual methods with the iterative approach has been performed. It was shown that only a combined approach using RBS together with PIXE, EBS and ToF-E ERDA yielded the total areal thickness of the alloy with inaccuracies of e.g. up to 8 % for RBS and beyond 10 % for an ERDA stand-alone analysis being observed. When determining stoichiometry the relative concentrations of the metal constituents Cr, Fe and Ni could be obtained with a higher precision, as in RBS and ERDA, using PIXE exclusively. The final accurate concentrations with improved precision required input in form of the matrix composition and thickness from both RBS and ERDA. The later indicated weak gradient of carbon in the alloy, ranging from ≈ 8 at.% close to the surface, up to ≈ 14 at.% nearby the $SiO_2$ layer. Combination with RBS could confirm the existence of this gradient as well as its relevance when, in turn, improving the RBS-fitting. Finally, the oxygen content present in the alloy, with a particular focus on the surface, was investigated by adopting the $^{16}O(\alpha, \alpha_0)^{16}O$ elastic reaction at 3.037 MeV to scan the bulk of the film. In comparison to ERDA, which also yields information surface oxygen, this method shows superior depth resolution close to the surface and more straightforward quantification independent from calibration.

The combination of several IBA techniques in an iterative and self-consistent analysis has proven to enhance the accuracy of the information that can be obtained from each independent measurement. In the present case, five different IBA spectra were analyzed simultaneously, yielding a remaining systematic uncertainty of the final description of the multi-layered sample in terms of its chemical composition depth-profile and thickness well below the average statistical accuracy, which is found better than ≈ 3 % in average.

**Acknowledgments**

Support by the Carl Tryggers foundation in form of a Postdoc scholarship. Radek Holeňák acknowledges financial support from the European Union program for education Erasmus. This study was performed in the framework of the competence center FunMat-II that is financially supported by Vinnova (grant no 2016-05156). Support by VR-RFI (contracts #821-2012-5144 & #2017-00646_9) and the Swedish Foundation for Strategic Research (SSF, contract RIF14-0053) supporting accelerator operation is gratefully acknowledged.

**References**

[1] J. R. Tesmer and M. Nastasi, *Handbook of Modern Ion Beam Materials Analysis*, 1st ed. (Materials Research Society, Warrendale, 1995).

[2] J. R. Bird, *Total analysis by IBA*, Nucl. Instrum. Methods B **45** (1990) 516-518

[3] J. P. Landsberg, *B. McDonald and F. Watt, Absence of aluminum in neurotic plaque cores in Alzheimer's disease*, Nature **360** (1992) 65-68.

[4] C. Jeynes, M. J. Bailey, N. J. Bright, M. E. Christopher, G. W. Grime, B. N. Jones, V. V. Palitsin and R. P. Webb, *"Total-IBA" - Where are we?,* Nucl. Instrum. Methods B **271** (2012) 107-118.

[5] W. K. Chu, J. M. Mayer, and M. A. Nicolet, *Backscattering spectrometry*, 1st ed. (Academic Press INC, San Diego, 1978).

[6] C.D.Lokhande, A. Ennaoui, P. S. Patil, M. Giersig, K. Diesner, M. Muller and H. Tributsch, *Chemical bath deposition of indium sulphide thin films: preparation and characterization*, Thin Solid Films **1-2** (1999) 18-23.

[7] E. Guziewicz, R. Ratajczak, M. Stachowicz, D. Snigurenko, T. A. Krajewski, C. Mieszczynski, K. Mazur, B. S. Witkowski, P. Dluzewski, K. Morawiec and A. Turos, *Atomic layer deposited ZnO films implanted with Yb: The influence of Yb location on optical and electrical properties*, Thin Solid Films **643** (2017) 7-15.

[8] C. Jeynes and J. Colaux, *Thin film depth profiling by ion beam analysis*, Analyst **141** (2016) 5944-5985

[9] J. R. Tesmer, C. J. Maggiore, M. Nastasi, J. C. Barbour and J. W. Mayer (eds.), *High energy and heavy ion beams in material analysis*, (Materials Research Society, Pittsburg, 1990).

[10] H. J. Whitlow, G. Possnert, C. S. Petersson, *Quantitative Mass and Energy Dispersive Elastic Recoil Spectrometry: resolution and efficiency considerations*, Nucl. Instrum. Methods B **27** (1987) 448-457.

[11] Hui-Ying Qu, D. Primetzhofer, M. A. Arvizu, Z. Qiu, U. Cindemir, C. G. Granqvist and Gunnar A. Niklasson, *Electrochemical Rejuvenation of Anodically Coloring Electrochromic Nickel Oxide Thin Films*, ACS Appl. Mater. Interfaces **9** (2017) 42420-42424.

[12] J. Jokinen, P. Haussalo, J. Keinonen, M. Ritala, D.Riihelä and M. Leskelä, *Analysis of AlN thin films by combining TOF-ERDA and NRB techniques*, Thin Solid Films **289** (1996) 159-165.

[13] R. A. Jarjis, *Nuclear cross section data for surface analysis*, Department of Physics, University of Manchester (1979).

[14] S. A. E. Johansson and J. L. Campbell, *PIXE: A novel technique for elemental analysis*, 1st ed. (John Wiley & Sons, New York 1988).

[15] J. N. Keuler, L. Lorenzen, R. D. Sanderson, V. Prozesky and W. J. Przybylowicz, *Characterization of electroless plated palladium–silver alloy membrane*, Thin Solid Films **347** (1999) 91-98.

[16] C. Jeynes, N. P. Barradas and E. Szilágyi, *Accurate Determination of Quantity of Material in Thin Films by Rutherford Backscattering Spectrometry*, Analytical Chemistry **84** (2012) 6061-6069.

[17] J. C. G. Jeynes, C. Jeynes, V. Palitsin and H. E. Townley, *Direct quantification of rare earth doped titania nanoparticles in individual human cells*, Nanotechnology **27** (2016) 285103-285111.

[18] D. B. Miracle, O. N. Senkov, A critical review of high entropy alloys and related concepts, Acta Materiala **122** (2017) 448-511.

[19] J. W. Yeh, S. K. Chen, S. J. Lin, J. Y. Gan, T. S. Chin, T. T. Shun, C. H. Tsau and S. Y. Chang, Nanostructured high-entropy alloys with multiple principal elements: novel alloy design concepts and outcomes, Adv. Engineering Mat. **6** (2004) 299-303.

[20] Y. F. Ye, Q. Wang, J. Lu, C. T. Liu and Y. Yang, High-entropy alloy: challenges and prospects, Materials Today **19** (2015) 349-362.

[21] H. J. Whitlow, G. Possnert and C. S. Petersson, Quantitative mass and energy dispersive elastic recoil spectrometry: resolution and efficiency considerations, Nucl. Instrum. Methods B **27** (1987) 448-457.

[22] J. A. Leavitt, L. C. McIntyre Jr., M. D. Ashbaugh, J. G. Oder, Z. Lin and B. Dezfouly-Arjomandy, Cross sections for 170.5 backscattering of 4He from oxygen for 4He energies between 1.8 and 5.0 MeV, Nucl. Instrum. Methods B **44** (1990) 260-265.

[23] A. F. Gurbich, *Evaluated differential cross-sections for IBA*, Nucl. Instrum. Methods B **268** (2010) 1703-1710.

[24] J. L. Colaux, G. Terwagne and C. Jeynes, *On the traceably accurate voltage calibration of electrostatic accelerators*, Nucl. Instrum. Methods B **349** (2015) 173-183.

[25] V. Paneta, M. Kokkoris, A. Lagoyannis, and K. Preketes-Sigalas, *Accurate accelerator energy calibration using selected resonances in proton elastic scattering and in (p,c) and (p,p0c) reactions*, Nucl. Instrum. Methods B **406**, (2017) 108-111.

[26] P. Ström, P. Petersson, M. Rubel, and G. Possnert, *A combined segmented anode gas ionization chamber and time-of-flight detector for heavy ion elastic recoil detection analysis*, Rev. Sci. Instrum. **87** (2016) 103303-103308.

[27] Y. Zhang, H.J. Whitlow, T. Winzell, I.F. Bubb, T. Sajavaara, K. Arstila and J. Keinonen, *Detection efficiency of time-of-flight energy elastic recoil detection analysis systems*, Nucl. Instrum. Methods B **149** (1999) 477–489.

[28] M. Mayer, W. Eckstein, H. Langhuth, F. Schiettekatte, and U. V. Toussaint, *Computer simulation of ion beam analysis: Possibilities and limitations*, Nucl. Instrum. Methods B **269**, (2011) 3006-3013.

[29] J. F. Ziegler, *SRIM The stopping and range of ions in matter*. Available at http://www.srim.org/. Accessed on July 2018.

[30] T. F. Silva, M. V .Moro, G. F. Trindade, N. Added, M. H. Tabacniks, R. J. Santos, P. L. Santana and J. R. R. Bortoleto, *Ion Beam Analysis of a-C:H films on alloy steel substrate*, Thin Solid Films **545** (2013) 171-175.

[31] J. L. Campbell, N. I. Boyd, N. Grassi, P. Bonnick and J. A. Maxwell, *The Guelph PIXE software package IV*, Nucl. Instrum. Methods B **268** (2010) 3356-3363.

[32] T. F. Silva, C. L. Rodrigues, M. Mayer, M. V. Moro, G. F. Trindade, F. R. Aguirre, N. Added, M. A. Rizzutto and M. H. Tabacniks, *MultiSIMNRA: A computational tool for self-consistent ion beam analysis using SIMNRA*, Nucl. Instrum. Methods B **371** (2016) 86-89.

[33] D. Abriola, N. P. Barradas, I. Bogdanovic-Radovic, M. Chiari, A. F. Gurbich, C. Jeynes, M. Kokkoris, M. Mayer, A. R. Ramos, L. Shi, I. Vickridge, *Development of a reference database for Ion Beam Analysis and future perspectives*, Nucl. Instrum. Methods B **269** (2011) 2972-2978.

[34] K. Arstila, J. Julin, M. I. Laitinen, J. Aalto, T. Konu, S. Kärkkäinen, S. Rahkonen, M. Raunio, J. Itkonen, J. -P. Santanen, T. Tuovinen, T. Sajavaara, *Potku - New analysis software for heavy ion elastic recoil detection analysis*, Nucl. Instrum. Methods B **331** (2014) 34-41.

[35] K. Kantre, V. Paneta and D. Primetzhofer, *Investigation of the energy loss of I in Au at energies below the Bragg peak*, Nucl. Instrum. Methods B **450** (2018) 37-42.

[36] Working Group of the Joint Committee for Guides in Metrology, *Evaluation of Measurement Data: Guide to the Expression of Uncertainty in Measurement* (2008).

[37] Working Group of the Joint Committee for Guides in Metrology, *Evaluation of Measurement Data: Supplement 2 to the Guide to the Expression of Uncertainty in Measurement: Extension to Any Number of Output Quantities* (2011).

[38] K. A. Sjöland, F. Munnik, and U. Wätjen, *Uncertainty budget for Ion Beam Analysis*, Nucl. Instrum. Methods B ***161-163*** (2000) 275-280.

[39] J. L. Colaux and C. Jeynes, High accuracy traceable Rutherford backscattering spectrometry of ion implanted samples, Analytical Methods **6** (2014) 120-129.

[40] H. Paul and D. Sánchez-Parcerisa, *A critical overview of recent stopping power programs for positive ions in solid elements,* Nucl. Instrum. Methods B **312** (2013) 110-117.

[41] M. V. Moro, T. F. Silva, A. Mangiarotti, Z. O. Guimarães-Filho, M. A. Rizzutto, N. Added and M. H. Tabacniks, *Traceable stopping cross sections of Al and Mo elemental targets for 0.9-3.6 MeV protons,* Phys. Rev. A **93** (2016) 022704-022721.

[42] M. V. Moro, B. Bruckner, P. L. Grande, M. H. Tabacniks, P. Bauer and D. Primetzhofer, *Stopping cross section of vanadium for H+ and He+ ions in a large energy interval deduced from backscattering spectra,* Nucl. Instrum. Methods B **424** (2018) 43-51.

[43] G. Lulli, E. Albertazzi, M. Bianconi, G. G. Bentini, R. Nipoti and R. Lotti, *Determination of He electronic energy loss in crystalline Si by Monte-Carlo simulation of Rutherford backscattering-channeling spectra*, Nucl. Instrum. Methods B **170** (2000) 1-9.